\documentclass[useAMS,usenatbib]{mn2e}

\usepackage{natbib}
\usepackage{dcolumn}
\usepackage{graphicx}

\newcommand{\physrep}{Phys.~Rep.}

\newcommand{\be}{\begin{equation}}
\newcommand{\ee}{\end{equation}}
\newcommand{\bea}{\begin{eqnarray}}
\newcommand{\eea}{\end{eqnarray}}
\title[A New Independent Limit on $\Lambda$ from Bending of Light by Clusters of Galaxies]{A New Independent Limit on the Cosmological Constant/Dark Energy from the Relativistic Bending of Light by Galaxies and Clusters of Galaxies}
\author[M. Ishak, W. Rindler, J. Dossett, J. Moldenhauer, C. Allison]{M. Ishak\thanks{E-mail:
mishak@utdallas.edu}, W. Rindler\thanks{F.R.A.S.}, J. Dossett, J. Moldenhauer, and C. Allison\\ \\
Department of Physics, The University of Texas at Dallas, Richardson, TX 75083, USA}

\begin{document}

\date{\today}

\pagerange{\pageref{firstpage}--\pageref{lastpage}} \pubyear{0000}

\maketitle

\label{firstpage}

\begin{abstract}
We derive new limits on the value of the cosmological constant, $\Lambda$, based on the Einstein bending of light by systems where the lens is a distant galaxy or a cluster of galaxies. We use an amended lens equation in which the contribution of $\Lambda$ to the Einstein deflection angle is taken into account and use observations of Einstein radii around several lens systems. We use in our calculations a Schwarzschild-de Sitter vacuole exactly matched into a Friedmann-Robertson-Walker background and show that a $\Lambda$-contribution term appears in the deflection angle within the lens equation. We find that the contribution of the $\Lambda$-term to the bending angle is larger than the second-order term for many lens systems. Using these observations of bending angles, we derive new limits on the value of $\Lambda$. These limits constitute the best observational upper bound on $\Lambda$ after cosmological constraints and are only two orders of magnitude away from the value determined by those cosmological constraints. 
\end{abstract}

\begin{keywords}
cosmology: theory -- gravitational lensing -- gravitation.
\end{keywords}

\section{Introduction}

Cosmic acceleration and the dark energy associated with it constitute one of the most important and challenging current problems in cosmology and all physics, see for example the reviews \cite{rev1,rev2,rev3,rev4,rev5,rev6,rev7,rev8,rev9} and references therein. The cosmological constant, $\Lambda$, is among the favored candidates responsible for this acceleration. Current constraints on $\Lambda$ are coming from cosmology, see e.g. \cite{obs1,obs2,obs3,obs4,obs5,obs6,obs7,obs8,obs9,obs10}, and it is important to obtain constraints or limits from other astrophysical observations. 

Very recently, the authors of reference \cite{RindlerAndIshak2007} demonstrated that, contrarily to previous claims (e.g. 
\cite{Islam,Freire,Kagramanova,Finelli,Sereno,Kerr}), when the geometry of the Schwarzschild-de Sitter spacetime is taken into account, the cosmological constant does contribute to the light-bending around a concentrated source and hence to the corresponding Einstein deflection angle. This result was confirmed in \cite{Lake2007,Sereno2007,Schucker2007}.

In this paper, we incorporate that result into the broadly used lens equation and then apply it to current observations of Einstein radii around distant galaxies and clusters of galaxies. Using observational data of a selected list of Einstein radii around clusters and galaxies, we show that the contribution of the cosmological constant to the bending angle can be larger than the second-order term of the Einstein bending angle. These new results allow us to put new independent upper bounds on the value of the cosmological constant based on the observations of the bending angle by galaxies and clusters of galaxies. 
The results provide an improvement of eight orders of magnitude on previous upper bounds on $\Lambda$ from planetary or stellar systems, see for example \cite{Sereno,Kagramanova}. Interestingly, these limits provide the best observational upper bound on $\Lambda$ after cosmological constraints and are only two orders of magnitude away from the value determined by those cosmological constraints.

\begin{figure*}
\begin{center}
\includegraphics[width=6.5in,height=2.9in,angle=0]{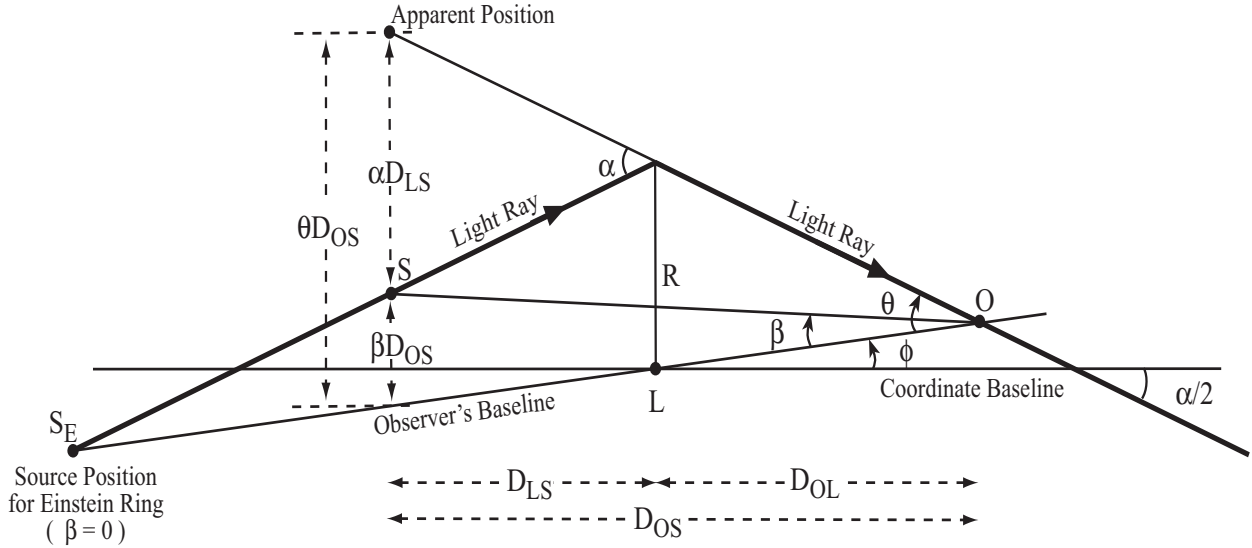}
\caption{\label{fig:figure1} 
The lens equation geometry. Observer, lens, and source are at O, L, and S, respectively. The position of the unlensed source is at an angle $\beta$, the apparent position is at the angle $\theta$ and the deflection angle is $\alpha$. The distance from the observer to the source is $D_{OS}$, from the observer to the lens is $D_{OL}$, and from the lens to the source is $D_{LS}$. The angle $\phi$ is as shown on the figure. As usual, the lens equation follows from the geometry as $\theta D_{OS}=\beta D_{OS} + \alpha D_{LS}$. 
}
\end{center}
\end{figure*}

\section[]{The bending angle in the presence of a cosmological constant}

We outline here the main steps of the calculation of \cite{RindlerAndIshak2007} and extend it using the second-order terms for the solution of the null geodesic equation. We consider the Schwarzschild-de Sitter (SdS) metric \cite{Kottler}
\begin{equation}
ds^2=f(r) dt^2 - f(r)^{-1} dr^2-r^2 (d\theta^2+sin^2(\theta) d\phi^2)
\label{eq:metric}
\end{equation}
where
\begin{equation}
f(r) \equiv 1-\frac{2m}{r}-\frac{\Lambda r^2}{3},
\label{eq:alpha}
\end{equation}
and where we use relativistic units ($c=G=1$), $m$ being the mass of the central object. 

As shown in many text books, e.g. \cite{Rindler,MTW}, the null geodesic equation in SdS spacetime is given exactly by  
\begin{equation}
\frac{d^2u}{d\phi^2}+u=3 m u^2, \,\,\,\, (u \equiv 1/r).
\label{eq:ODE}
\end{equation} 

In the usual way, the null orbit is obtained as a perturbation of the undeflected line 
(i.e. the solution of equation (\ref{eq:ODE}) without the RHS)
\begin{equation}
r \, sin(\phi)= R.
\label{eq:straightline}
\end{equation}
After substitution of (\ref{eq:straightline}) into (\ref{eq:ODE}), one obtains the following equation for $u$ (Eq. (11.64) in \cite{Rindler}) 
\begin{equation}
\frac{1}{r}=u=\frac{sin(\phi)}{R}+\frac{3m}{2R^2}\Big{(}1+\frac{1}{3}cos(2\phi)\Big{)},
\label{eq:solution}
\end{equation}
where $R$ is a constant related to the physically meaningful area distance $r_{0}$ of closest approach (when $\phi=\pi/2$) by 
\begin{equation}
\frac{1}{r_{0}}=\frac{1}{R}+\frac{m}{R^2}.
\label{eq:r_0R}
\end{equation}
Many authors, see for example \cite{Wald1984,MTW}, use the impact parameter $b$ to discuss the bending of light in Schwarzschild spacetime,  but SdS spacetime is not asymptotically flat and one needs to define another parameter such as $R$. As shown in \cite{RindlerAndIshak2007}, the contribution of $\Lambda$ to the bending angle comes from the spacetime metric itself, independently of the parameterization of the null geodesic equation.

It was shown in \cite{RindlerAndIshak2007} that the angle $\theta$ of our Figure 1 (denoted by $\psi$ in \cite{RindlerAndIshak2007}) is given by 
\begin{equation}
\tan(\theta)=\frac{f(r)^{1/2}r}{|dr/d\phi|}.
\label{eq:tan}
\end{equation}
with $f(r)$ as in Eq.(\ref{eq:alpha}) above ($f(r)$ is $\alpha(r)$ in \cite{RindlerAndIshak2007}) and 
\begin{equation}
\frac{dr}{d\phi}=\frac{m r^2}{R^2} \sin (2 \phi)-\frac{r^2}{R} \cos (\phi)
\label{eq:A}
\end{equation}
to lowest order. The total bending angle $\alpha$ (at coordinate $\phi=0$, just so as to have some standard position at which to measure it) was found in \cite{RindlerAndIshak2007} to be 
\begin{equation}
\alpha \approx 4 \frac {m}{R} - \frac {\Lambda{R}^{3}}{6 m}. 
\label{eq:result1}
\end{equation}
to first order in $m/R$. This result shows that a positive $\Lambda$ diminishes $\alpha$, as might well be expected from the repulsive effect of $\Lambda$. The first term in (\ref{eq:result1}) is simply the classical Einstein bending angle to first order. 

Now, since we plan to compare to observations, it is useful to expand the calculation to higher orders   including the second-order solution to the null geodesic equation. In the usual way, see for example \cite{Bodenner}, we write
\begin{equation}
u=u_0 [sin(\phi) + (m u_0) \delta u_1 + (m u_0)^2 \delta u_2]
\label{eq:SecondOrderTrial} 
\end{equation}
where $u \equiv \frac{1}{r}$ and $u_0 \equiv \frac{1}{R}$.
Substituting this into equation (\ref{eq:ODE}) and collecting terms of equal powers of $Mu_0$ gives the following two equations: 
\begin{equation}
\frac{d^2 \delta u_1}{d\phi^2}+\delta u_1 = 3 \sin^2 \phi
\label{eq:ODE1}
\end{equation}
\begin{equation}
\frac{d^2 \delta u_2}{d\phi^2}+\delta u_2 = 6 \delta u_1 \sin \phi.
\label{eq:ODE2}
\end{equation}
Solving (\ref{eq:ODE1}) and (\ref{eq:ODE2}) for $\delta u_1$ and $\delta u_2$ and substituting them into (\ref{eq:SecondOrderTrial}) gives the solution 
\begin{eqnarray}
\frac{1}{r}&=&\frac{\sin \phi}{R} + \frac{3 m}{2 R^2}\Big{(}1+\frac{\cos 2 \phi}{3} \Big{)} +  \frac{3 m^2}{16 R^3}\Big{(}10 \pi \cos \phi - \nonumber \\& & \,\,\,\,20\phi \cos \phi - \sin 3 \phi\Big{)}.
\label{eq:SecondOrderSolution} 
\end{eqnarray}

Now, we differentiate (\ref{eq:SecondOrderSolution}) and multiply by $r^2$ to obtain 
\begin{eqnarray}
\frac{dr}{d\phi}&=&-\frac{r^2}{R} \cos \phi + \frac{m r^2}{R^2} \sin 2\phi+ \frac{15m^2 r^2}{4 R^3}\Big{(}\cos \phi +
\nonumber \\
&  & \,\,\,\,\,\,\, \frac{3}{20} \cos 3\phi + (\frac{\pi}{2}-\phi)\sin \phi\Big{)}.
\label{eq:A2}
\end{eqnarray}
After some manipulation, it follows from (\ref{eq:tan}) and (\ref{eq:A2}) that the total bending angle (at $\phi=0$) to the third-order is given by 
\begin{equation}
\alpha \approx 4 \frac {m}{R}+ \frac{15 \pi}{4}\frac {{m}^{2}}{{R}^{2}} + \frac {305}{12} \frac {{m}^{3}}{{R}^{3}}-
\frac {\Lambda{R}^{3}}{6 m}. 
\label{eq:result2}
\end{equation}
The coefficients for the first and second-order terms in this expansion are the same as the ones in the expansion in terms of the impact parameter $b$, see e.g. \cite{Keeton}, that is used for the asymptotically flat Schwarzschild spacetime. In the next section, we put our results into an observational context using systems where the lens is a galaxy or a cluster of galaxies.

\section{Observations of Einstein-Radii and the contribution of the Cosmological Constant to the deflection}
As one might expect, while the cosmological constant has a very negligible effect on small scales this is not the case at the level of galaxies and clusters of galaxies. In this section, we evaluate the contribution of the cosmological constant to the bending of light using observations of large Einstein radii where the lens is a galaxy or a cluster of galaxies.   

Equations (\ref{eq:result1}) and (\ref{eq:result2}) above were derived based on a source and an observer located in a Schwarzschild-de Sitter background. We will derive here the corresponding equation in a Friedmann-Lemaitre-Robertson-Walker background (FLRW). For that, we consider a Schwarzschild-de Sitter vacuole exactly embedded into an FLRW spacetime using the Israel-Darmois formalism \cite{Darmois,Israel}. The relations between radial coordinates $r_{b}$ at the boundary of the vacuole are simple and well-known in the literature, see for example \cite{Swiss-cheese,Swiss-cheese2}, and are given by the following two equations:
\be
r_{b\,\, in \, SdS} = a(t) \,\,\textit{r}_{b\,\,in\,FLRW}
\label{eq:cond1}
\ee
and
\be
m_{\,SdS} = \frac{4 \pi}{3}\,\, r_{b\,\, in \, SdS}^3\times \rho_{matter\,in\,FLRW}.        
\label{eq:cond2}
\ee
Thus, for a given cluster mass, equation (\ref{eq:cond2}) provides a boundary radius where the spacetime transitions from a SdS spacetime to an FLRW background. We shall assume that all the light-bending occurs in the SdS vacuole according to our previous formulae, and that once the light transitions out of the vacuole and into FLRW spacetime, all $\Lambda$-bending stops. Unlike the mass-effect,  which falls off quickly, the $\Lambda$-effect on the bending of light increases with distance from the source (the "$\Lambda$-repulsion" is proportional to distance);  hence the question of where to cut off the integration becomes important.  The choice of the boundary of the vacuole ($r_b$)  in the Einstein-Strauss model seems physically the most appropriate, whereas the choice $\phi=0$ in ref. \cite{RindlerAndIshak2007} was purely conventional.

Now, for the small angle $\phi_b$ at the boundary, equation (\ref{eq:solution}) gives 
\be
u_b=\frac{1}{r_b}=\frac{\phi_b}{R}+\frac{2m}{R^2}
\label{eq:u_boundary}
\ee
and equation (\ref{eq:A}) gives 
\be
|A|=\frac{r_b^2}{R}\Big{(}1-\frac{2\phi_b m}{R}\Big{)}.
\label{eq:A_boundary}
\ee
Next, inserting (\ref{eq:u_boundary}) and (\ref{eq:A_boundary}) into equation (\ref{eq:tan}) yields after a few steps 
\be
\theta \approx \tan{\theta} \approx \phi_b + \frac{2m}{R}-\frac{\Lambda \phi_b r_b^2}{6} + \,\,\textrm{higher-order terms}.
\ee
The bending angle, $\alpha$, is given, to the smallest order in $m/R$ and $\Lambda$, by 
\be
\frac{\alpha}{2}\approx \theta - \phi_b \approx \frac{2m}{R}-\frac{\Lambda \phi_b r_b^2}{6}.
\label{eq:epsilon}
\ee
Now, equation (\ref{eq:u_boundary}) yields, to the smallest order, $\phi_b=R/r_b$, so we can finally write from (\ref{eq:epsilon})
\be
\alpha\approx\frac{4m}{R}-\frac{\Lambda R r_b}{3}
\ee
where $R$ is related to the closest approach by equation (\ref{eq:r_0R}) and $r_b$ is the boundary radius between SdS and FLRW, and is given by equation (\ref{eq:cond2}). 

Perhaps a caveat that one need to address is that the Einstein-Strauss model that was used here is known to have some instability to radial perturbations at the boundary as, for example, discussed in \cite{Krasinski} and references therein. However, our work hinges on finding a cut-off location where the $\Lambda$-bending of the lens can be regarded as accomplished. In the predecessor paper to this one \cite{RindlerAndIshak2007} we chose  $\phi = 0$  as the only readily available standard cut-off point. The present paper is an improvement over the previous one in this respect, in that we now have a cut-off point tailored to each individual lens, namely the edge of the vacuole. The vacuole model as such is not used except for this one purpose, namely to give us a realistic order-of-magnitude estimate of the "range of influence" of the lens. Moreover, as it is widely used in gravitational lensing studies, one could also resort to approximation methods where the inhomogeneity is modeled by a gravitational potential that is embedded in an FLRW background \cite{Bartelmann,Mellier,Carroll}. Such an alternative treatment of the questions addressed here has been recently carried out in \cite{Ishak2008} and has confirmed the findings of the present work. 

Also, our result is expressed in terms of the vacuole boundary $r_b$ that is evaluated at some instant in time. The vacuole and its boundary expand as the universe expands thus when we calculate $r_b$ from equation (\ref{eq:cond2}) we must use the density of the universe as it was when light passed by the lens. We are aware of the instability of $r_b$ but since we need it at one instant, the instability should not affect our result.

Next, using equations (\ref{eq:SecondOrderSolution}) and (\ref{eq:A2}), we can expand the result to 
\be
\alpha \approx 4 \frac {m}{R}+ \frac{15 \pi}{4}\frac {{m}^{2}}{{R}^{2}} + \frac {305}{12} \frac {{m}^{3}}{{R}^{3}} -\frac{\Lambda R r_b}{3}
\label{eq:alpha_final}
\ee

\begin{table*}   
\begin{center}
\caption{Contributions of the cosmological constant to the Einstein bending angle by distant clusters of galaxies. Column-8 shows that the $\Lambda$-term contribution is larger than the second-order term in the Einstein bending angle for these lens systems. The last column shows limits on the cosmological constant based on observations of the bending angle. These limits provide the best upper bound on $\Lambda$ after cosmological constraints and are only two orders of magnitude away from the value determined for $\Lambda$ by those cosmological constraints, i.e. $1.29\,\,10^{-56} cm^{-2}$. Previously, the best upper bound after cosmology was determined from planetary or stellar systems and is $\Lambda \leq  \,\,10^{-46} cm^{-2}$, see for example (Sereno \& Jetzer 2006, Kagramanova et al. 2006) and references therein.}
{\small
\begin{tabular}{ccccccccc}
\hline
Cluster or galaxy&Einstein&Mass in&1st Order&2nd Order&$\Lambda$-term& Ratio-1&Ratio-2&Upper Limit\\
name and references&Radius&$M_{sun}h^{-1}$&term&term&(rads)&1st/$\Lambda$-term& $\Lambda$-term/2nd &on $\Lambda$\\
 & (Kpc) & & (rads)& (rads)& & & &($cm^{-2}$)\\
\hline
\hline
Abell 2744 & 96.4 &$1.97\times 10^{13}$& 5.53E-05 & 2.25E-09 & 1.68E-08 & 3.28E+03 & 7.48 & 4.23E-54\\
\cite{Smail1991,Allen}&&&&&&&&\\
\hline
Abell 1689 & 138.2 &$9.36\times 10^{13}$& 1.88E-04 & 2.61E-08 & 4.52E-08 & 3.52E+03 & 1.73 & 5.37E-54\\
\cite{Allen,Limousin}&&&&&&&&\\
\hline
SDSS J1004+4112& 110.0 &$4.26\times 10^{13}$& 1.05E-04 & 8.06E-09 & 2.22E-08 & 4.70E+03 & 2.76 & 6.07E-54\\
 \cite{Sharon}&&&&&&&&\\
\hline
3C 295 & 127.7 &$7.1\times 10^{13}$& 1.50E-04 & 1.66E-08 & 3.06E-08 & 4.90E+03 & 1.84 & 6.33E-54\\ 
\cite{Wold}&&&&&&&&\\
\hline
Abell 2219L & 86.3 &$3.22\times 10^{13}$& 1.01E-04 & 7.47E-09 & 1.85E-08 & 5.44E+03 & 2.48 & 7.01E-54\\
\cite{Smail1995a,Allen}&&&&&&&&\\
\hline
AC 114 & 54.6 &$9.23\times 10^{12}$& 4.57E-05 & 1.54E-09 & 7.38E-09 & 6.19E+03 & 4.80 & 7.99E-54\\
\cite{Smail1995b,Allen}&&&&&&&&\\ 
\hline
\end{tabular} }
\end{center}
\end{table*}

Finally, following the usual procedure,  see e.g. \cite{Mellier,Bartelmann}, we put our results into the lens equation which is given from the geometry (see Figure 1) and small-angle relations as follows  
\be
\theta D_{OS}=\beta D_{OS} + \alpha D_{LS}  
\ee
or in the familiar form
\be
\theta=\beta+\alpha \,\frac{D_{LS}}{D_{OS}}
\label{eq:lensequation}
\ee
where all the quantities are as defined in Figure 1, and the angular-diameter distance is given by 
\be
D(z)=\frac{c}{H_0(1+z)}\int^{z}_{0}\frac{dz'}{\sqrt{\Omega_m(1+z')^3+\Omega_\Lambda}}
\ee
where, for the spatially flat concordance cosmology, $\Omega_m=0.27$, $\Omega_\Lambda=0.73$, and $H_0=71 km/s/Mpc$. 

Thanks to the advancement of observational techniques, one can find in the literature a number of distant galaxies and clusters of galaxies that are lenses with large Einstein radii, making them very interesting for applying our results. The selected systems are shown in Table 1 along with our evaluation of the deflection first-order term, the second-order term, and the $\Lambda$-term, and some of their ratios. Despite the smallness of the cosmological constant, $\Lambda$, we find that the Einstein first-order term in the bending 

angle due to these systems is only by some $10^3$ bigger than the $\Lambda$-term. Interestingly, we find that for the lens systems in Table 1, the contribution of the cosmological constant term is larger than the second-order term of the Einstein bending angle.\\

\section{A new limit on the Cosmological Constant from Light-Bending}
From cosmology (e.g. using supernova magnitude-redshift relation and the Cosmic Microwave Background Radiation), the value of the cosmological constant, $\Lambda$, is found to be about $1.29\,\,10^{-56} cm^{-2}$ (using $H_0=71$ km/s/Mpc and $\Omega_{\Lambda}=0.73$, see e.g. \cite{RindlerLambda,obs1,obs2,obs3,obs4,obs5,obs6,obs7,obs8,obs9,obs10}). It is very desirable to obtain other limits on  $\Lambda$ that come from other astrophysical constraints. 
As we show, when we consider the uncertainty in the measurements of the bending angle (which is around $\Delta \alpha \sim$ 5-10$\%$ for several of the systems considered in Table 1), we find that the bending angle due to distant galaxies and clusters can provide interesting limits on the value of the cosmological constant. Indeed, if the contribution of $\Lambda$ cannot exceed the uncertainty in the bending angle for these system, then it follows that 
\begin{equation}
\Lambda \leq \frac{3\, \Delta \alpha}{R\,\, r_b}.  
\end{equation}
For example, with $\Delta \alpha = 10\%$, we find from the system Abell 2744 \cite{Smail1991,Allen} that   
\be
\Lambda \leq  4.23\,\,\,10^{-54} cm^{-2}.
\ee
The other limits are in Table 1. Interestingly, these limits are the best observational upper bound on the value of $\Lambda$ after cosmological constraints and are only two orders of magnitude away from the value determined from cosmological constraints. In fact, $\Lambda$ also enters into the expression of the angular diameter distance but our estimation is that it can affect our limit by a factor of two or less. Previously, the best upper bound after cosmology was provided from planetary or stellar systems and is $\Lambda \leq  \,\,10^{-46} cm^{-2}$, see for example \cite{Sereno,Kagramanova} and references therein. 

\section{Conclusion}
In conclusion, we showed that a $\Lambda$-contribution term appears in the deflection angle within the lens equation. This contribution is larger than the second-order term in the Einstein bending angle for many cluster lens systems. These results allow us to put new upper bounds on the cosmological constant, $\Lambda$, based on observations of the bending angle by galaxies and clusters of galaxies. These results provide an improvement of 8 orders of magnitude on previous upper bounds on $\Lambda$ that were based on planetary or stellar systems, e.g. \cite{Sereno,Kagramanova}. The limits provide the best upper bound on $\Lambda$ after cosmological constraints and are only two orders of magnitude away from the value determined for $\Lambda$ from those cosmological constraints. 


\section*{Acknowledgments}
The authors thank C. Kochanek and I. Browne for useful comments about some lens systems. M.I. acknowledges partial support from the Hoblitzelle Foundation.


\label{lastpage}


\begin{thebibliography}{99}
%
\bibitem[Albrecht et al. 2006]{rev8} Albrecht A. et al, {\it Report of the Dark Energy Task Force} astro-ph/0609591 (2006).
\bibitem[Allen 1998]{Allen} Allen S., M.N.R.A.S, \textbf{296}, 392, (1998).
\bibitem[Bartelmann \& Schneider 2001]{Bartelmann}Bartelmann M., Schneider P. Phys.Rept. 340, 291 (2001).
\bibitem[Bennett et al. 2003]{obs5} Bennett C., {\em{et al.}}, Astrophys. J. Suppl. Ser. {\textbf{148}}, 1 (2003).
\bibitem[Bodenner \& Will] {Bodenner} Bodenner J., Will C., Am. J. Phys., Vol. {\textbf{71}}, No. 8, 770 (2003).
\bibitem[Carroll 2001]{rev4} Carroll S., Living Reviews in Relativity, \textbf{4}, 1 (2001).
\bibitem[Carroll 2004]{Carroll} Carroll S., \textit{Spacetime and Geometry: An Introduction to General Relativity} (Addison Wesley, San Fransisco, 2004).
\bibitem[Darmois 1927]{Darmois} Darmois G.{\em M\'{e}morial de Sciences Math\'{e}matiques, Fascicule XXV}, ``Les equations de la gravitation einsteinienne'', Chapitre V. (1927).
\bibitem[Einstein \& Strauss 1945]{Swiss-cheese} Einstein A. and Strauss E., Rev. Mod. Phys. \textbf{17}, 120.(1945), erratum: ibid {\bf 18}, 148 (1946). 
\bibitem[Finelli et al. 2007]{Finelli} Finelli F., Galaverni M., Gruppuso A., Phys.Rev. D {\textbf{75}}, 043003 (2007).
\bibitem[Freire et al. 2001]{Freire} Freire W., Bezerra V.,  Lima J., Gen. Relativ. Gravit. {\textbf{33}}, 1407 (2001).
\bibitem[Ishak 2007] {rev9}Ishak M., Foundations of Physics Journal, Vol.\textbf{37}, No 10, 1470 (2007).  
\bibitem[Ishak 2008] {Ishak2008}Ishak M., arXiv:0801.3514v1 [astro-ph].
\bibitem[Islam 1983]{Islam} Islam N., Phys. Lett. A  {\textbf{97}}, 239 (1983).
\bibitem[Israel 1966]{Israel} Israel W., Nuovo Cim B {\bf 44}, 1. (1966). Erratum {\bf 48}, 463. 
\bibitem[Kagramanova et al. 2006]{Kagramanova}  Kagramanova V., Kunz J., Lammerzahl C., Phys.Lett. B {\textbf{634}} 465-470 (2006).
\bibitem[Keeton \& Petters]{Keeton} Keeton C., Petters A., Phys. Rev. D{\textbf{72}}, 104006 (2005).
\bibitem[Kerr et al. 2003]{Kerr} Kerr A., Hauck J., Mashhoon B.,  Class. Quant. Grav., {\textbf{20}}, 2727 (2003)). However, we clarify here that this reference (also cited in our previous paper \cite{RindlerAndIshak2007}) was rather focused on the study of the influence of $\Lambda$ on the time delay and not light deflection angle. The statement there about $\Lambda$ and its influence on light deflection was based on other previous work.  
\bibitem[Knop et al. 2003]{obs3} Knop R., {\em{et al.}}, Astrophys. J. {\textbf{598}}, 102-137 (2003).
\bibitem[Kottler 1918]{Kottler} Kottler F., Ann. Phys. {\textbf{361}}, 401 (1918).
\bibitem[Krasinski 1997]{Krasinski} Krasinski A., \textit{Inhomogeneous Cosmological Models} (Cambridge University Press, Cambridge, 1997). 
\bibitem[Lake 2007]{Lake2007}Lake K., arXiv:0711.0673.
\bibitem[Limousin 2007]{Limousin} Limousin M. et al.  Astrophys. Jour., 668:643–666, (2007). 
\bibitem[Mellier 1999]{Mellier} Mellier Y., Ann.~Rev.~Astron.~Astrophys. {\textbf{37}},  127 (1999).
\bibitem[Miralda-Escude \& Babul 1995]{Escude} Miralda-Escude J., Babul A., Astrophys. Jour., Vol. \textbf{449}, 18, (1995).
\bibitem[Misner Thorne and Wheeler 1973]{MTW} Misner C., Thorne K. and Wheeler J., \textit{Gravitation} (Freeman, San Francisco 1973).
\bibitem[Padmanabhan 2003]{rev5} Padmanabhan T., \physrep, \textbf{380}, 235 (2003).
\bibitem[Page et al. 2003]{obs7} Page L. et al., Astrophys. J. Suppl. Ser. {\textbf{148}}, 2333 (2003).
\bibitem[Peebles \& Ratra 2003]{rev6} Peebles J. and Ratra B., Rev. Mod. Phys. \textbf{75}, 559 (2003).
\bibitem[Perlmutter et al. 1999]{obs2} Perlmutter S., {\em{et al.}}, Astrophys. J. {\textbf{517}}, 565-586 (1999).
\bibitem[Riess et al. 1998]{obs1} Riess A., {\em{et al.}}, Astron. J. {\textbf{116}}, 1009-1038 (1998).
\bibitem[Riess et al. 2004]{obs4} Riess A., {\em{et al.}}, Astrophys. J. {\textbf{607}}, 665-687 (2004).
\bibitem[Rindler \& Ishak 2007]{RindlerAndIshak2007} Rindler W., Ishak M., Phys. Rev. D \textbf{76} 043006 (2007).
\bibitem[Rindler 1969]{RindlerLambda} Rindler W., Astrophys. J. \textbf{157}, L147 (1969). 
\bibitem[Rindler 2006]{Rindler} Rindler W., \textit{Relativity: Special, General, and Cosmological, Second Edition} (Oxford University Press, 2006).
\bibitem[Sahni \& Starobinsky 2000]{rev3} Sahni V., Starobinsky A. Int.J.Mod.Phys. \textbf{D9}, 373 (2000)
\bibitem[Schucker 2007]{Schucker2007}Schucker T., arXiv:0712.1559v1 [astro-ph].
\bibitem[Schucking 1954]{Swiss-cheese2} Schucking E., Z. Phys. 137, 595 (1954); and a long list of references in \cite{Krasinski} 
\bibitem[Seljak et al. 2005]{obs8} Seljak U. et al., Phys.Rev. D{\textbf{71}}, 103515 (2005).
\bibitem[Sereno \& Jetzer 2006]{Sereno} Sereno M., Jetzer Ph., Phys. Rev. D {\textbf{73}}, 063004 (2006).
\bibitem[Sereno 2007]{Sereno2007} Sereno M., arXiv:0711.1802.
\bibitem[Sharon 2006]{Sharon} Sharon K. et al., Astrophys. Jour. 629, L73 (2005); N. Ota, et al., Astrophys. Jour., 647, 215 (2006).
\bibitem[Smail et al. 1991]{Smail1991} Smail I. et al., M.N.R.A.S, 252, 19 (1991).
\bibitem[Smail et al. 1995a]{Smail1995a} Smail I., et al., MNRAS, 277, 1 (1995).
\bibitem[Smail et al. 1995b]{Smail1995b} Smail I., Couch W., Ellis R., Sharples R., Astrophys. Jour., 440, 501 (1995).
\bibitem[Spergel et al. 2003]{obs6} Spergel D., {\em{et al.}}, Astrophys. J. Suppl. Ser. , 175 (2003).
\bibitem[Spergel et al. 2007]{obs10} Spergel D., {\em{et al.}}, Astrophys. J. Suppl. 170, 377 (2007).
\bibitem[Tegmark et al. 2004]{obs9} Tegmark M., {\em{et al.}},  Astrophys. J. {\textbf{606}}, 702-740 (2004).
\bibitem[Turner 2000]{rev2} Turner M. ,\physrep, \textbf{333}, 619 (2000).
\bibitem[Upadhye et al. 2005]{rev7}Upadhye A., Ishak M., Steinhardt P., Phys. Rev. D \textbf{72}, 063501 (2005).
\bibitem[Wald 1984]{Wald1984} Wald R. \textit{General Relativity} (University of Chicago Press, 1984).
\bibitem[Weinberg 1989]{rev1} Weinberg S., Rev. Mod. Phys., \textbf{61}, 1 (1989).
\bibitem[Wold et al. 2002] {Wold} Wold M. et al. Mon.Not.Roy.Astron.Soc. 335, 1017 (2002).

\end{thebibliography}
\end{document}